# Casimir Terms and Shape Instabilities for Two-Dimensional Critical Systems

*Peter Kleban\**

Laboratory for Surface Science and Technology
& Department of Physics and Astronomy
University of Maine
Orono ME 04469, USA

*and*

*Ingo Peschel\*\**

Fachbereich Physik
Freie Universität Berlin
Arnimallee 14
D-14195 Berlin
Germany

**Abstract:** We calculate the universal part of the free energy of certain finite two-dimensional regions at criticality by use of conformal field theory. Two geometries are considered: a section of a circle ("pie slice") of angle $\phi$ and a helical staircase of finite angular (and radial) extent. We derive some consequences for certain matrix elements of the transfer matrix and corner transfer matrix. We examine the total free energy, including non-universal edge free energy terms, in both cases. A new, general, *Casimir instability* toward sharp corners on the boundary is found; other new instability behavior is investigated. We show that at constant area and edge length, the rectangle is unstable against small curvature.

\*e-mail: kleban@maine.maine.edu
\*\*e–mail: peschel@aster.physik.fu-berlin.de

**1. Introduction**

The free energy of a finite two-dimensional region at a critical point includes a universal term $F$. When the critical fluctuations are conformally invariant, this term may be calculated [1, 2, 3, 4, 5] via conformal field theory [6], without reference to any particular physical realization of the symmetry. It constitutes an example of the Casimir effect, first encountered in quantum electrodynamics [7].

Such terms have already been calculated in several geometries. In an infinite strip of width L, with identical boundary conditions on either side, $F$ is proportional to $-1/L$ [2, 8], which may be understood as an attraction of the walls. This attraction is also apparently present in a rectangular geometry [4], where it can lead to a thermodynamic force favoring the *elongation* [9] of very small real systems. However, we are not aware of any general (or even heuristic) argument for the qualitative behavior of such Casimir terms. Thus it is of interest to examine new shapes, in the hope of a more general understanding of the characteristic behavior of $F$.

For regions at two-dimensional phase transitions, the systems of interest here, $F$ includes two types of terms. There are both *scale-invariant* contributions, i.e. terms independent of the size of the system, but varying with its shape, and in addition *trace anomaly* terms which are *logarithmic* in the size. The latter will include contributions from any curved parts of the region's boundaries, curved metric, or corners on the boundary [3].

Our results, presented below, are valid for any conformally invariant (uniform) boundary condition. There will be, however, a geometry-independent term in $F$ that within a given universality class may change according to the specific boundary condition, or be independent of it [10]. Such contributions appear as integration constants in the procedure, described below, that we employ to calculate $F$, and have been suppressed in the presentation of our results.

For the circular section, considered in Section II, we calculate $F$ in what has become the standard manner. Derivatives of $F$ with respect to parameters of the geometry are determined by integrals of the expectation value of the stress tensor $\langle T \rangle$. This is, in turn, given by the Schwarzian derivative of the transformation from the half-plane to the geometry in question. The derivatives may then be integrated to determine $F$ up to an integration constant. The logarithmic terms are not necessarily included in this procedure, however they can be included in a straightforward way using the results of [3].

For the helical geometry, examined in Section III, $F$ is obtained via a transformation of the corresponding term in a rectangle [4], using the transfer matrix representation.

Section IV contains a derivation of some results for certain matrix elements of the transfer matrix and corner transfer matrix.

In Section V we examine the total free energy, including non-universal edge free energy terms, in both geometries. For the circular section, a new form of *Casimir instability*, favoring small angles (sharp corners) is investigated. The staircase is shown to have a similar instability, but one in which the edge length remains fixed, so there is no competition with the edge free energy term. We argue that such an instability is to be expected whenever sharp corners can form on a boundary. In another limit the competition between edge and Casimir terms is independent of area, by contrast to the usual situation. Finally, we show that at constant area and edge length, the rectangle is unstable against small curvature.

All free energies in this article are in units of $k_B T_c$, so that any $F$ is in fact dimensionless.

**II. Circular Section**

Here we consider a circular section of radius R and angle $\phi$, a "pie slice". In polar coordinates this is the region ($0 \leq r \leq R$, $0 \leq \theta \leq \phi$) in the w–plane. The upper half-plane z is mapped onto this geometry via the transformation

$$z = \left( \frac{1 + [w/R]^\nu}{1 - [w/R]^\nu} \right)^2, \tag{1}$$

where $\nu = \pi/\phi$, $\nu \geq 1/2$. Making use of the fact that $<T(z)> = 0$, we have [6]

$$<T(w)> = \frac{c}{12}\{z,w\}, \tag{2}$$

where c is the central charge and the curly brackets denote the Schwarzian derivative. We next decompose (1) into three successive transformations, $z = \zeta^2$, $\zeta = (1+\xi)/(1-\xi)$, and $\xi = (w/R)^\nu$. Effective use may then be made of the composition rule for Schwarzian derivatives

$$\{z_1,z_2\}(dz_2)^2 = \{z_1,z_3\}(dz_3)^2 + \{z_3,z_2\}(dz_2)^2. \tag{3}$$

After some computation, one finds

$$<T(w)> = -\frac{c}{2}\nu^2 \frac{u^{2\nu-2}}{(1-u^{2\nu})^2} + \frac{c}{24}(1-\nu^2)\frac{1}{u^2}, \tag{4}$$

where $u = w/R$. Note that the second term in (4) is exactly the stress tensor for an infinite wedge of angle $\phi$ [3], except that $|u| \leq 1$ in the present geometry. The first term, up to a factor of w, is proportional to a derivative. It is straightforward to verify that (4) has the correct divergence, with coefficient determined by the interior angle, as w approaches any of the three corners [3].

Now the universal term in the free energy $F$ can only depend on R and $\phi$. We next compute derivatives of $F$ with respect to these two parameters by use of the general formula [11]

$$\delta F = \frac{1}{2\pi}\int d^2r <T_{ij}> \frac{d\alpha^j}{dr_i}, \tag{5}$$

which gives the change in $F$ induced by a general coordinate transformation $z \to z + \alpha(z)$.

We make use of (5) to compute the derivative of $F$ with respect to $\phi$ by applying, in radial coordinates, the angular stretching transformation $r \to r$, $\theta \to \theta + \delta\phi\,\Theta(\theta-\theta_0)$, where $\Theta$ is the unit step function and $0 \leq \theta_0 \leq \phi$. One finds, using equation (2.2) in [3]

$$\delta F = \frac{1}{2\pi}\int <T_{\theta\theta}> r\,dr\,\delta\phi, \tag{6}$$

where $T_{\theta\theta} = -(e^{2i\theta}T + e^{-2i\theta}\bar{T})$, and the integrand is evaluated at $\theta = \theta_0$. Using (4) then leads to the expression

$$\delta F = \frac{c}{24\pi}\left(\left(\frac{\pi}{\phi}\right)^2 - 1\right)\ln(\frac{R}{r_c})\delta\phi - \frac{c}{8\phi}\delta\phi, \tag{7}$$

where $r_c$ is a short–distance cutoff that must be introduced to keep the integration over r finite. The two terms on the r.h.s. of (7) correspond to the two terms on the r.h.s. of (4). Note that the final expression is independent of $\theta_0$, as it should be.

The next step considers a radial stretch, $r \to r + \delta R\, \Theta(r-r_0)$, $\theta \to \theta$, where $0 \leq r_0 \leq R$. Making use of equation (2.13) in [3] leads to

$$\delta F = -\frac{1}{2\pi}\int <T_{rr}> \frac{1}{r}\Theta(r-r_0)d^2r\,\delta R + \frac{1}{2\pi}\int <T_{rr}> r_0 d\theta\,\delta R, \tag{8}$$

where the integrand in the second term is evaluated at $r = r_0$, and $T_{rr} = -T_{\theta\theta}$. The latter assumes that the trace $\Theta$ of the stress tensor vanishes, which is in fact not the case along the curved boundary and at the corners of the region [3]. Thus the logarithmic (trace anomaly) terms in $F$ are not properly accounted for, an omission that will be corrected below. Substituting (4) and performing the integrals gives

$$\delta F = \frac{c}{24}\left(\frac{\phi}{\pi} - \frac{\pi}{\phi}\right)\frac{\delta R}{R}, \tag{9}$$

which arises entirely from the second (infinite wedge) term in (4).

On calculating $\delta^2 F / \delta R \delta \phi$ from (7) and (9), one does not obtain the same expression. The difference, subtracting the latter from the former, is $-c/12\pi R$. However, according to the arguments in [3], (7) will contain a contribution from the curved part of the boundary, while (9) will not. The corresponding term in $F$ is $-(c\phi/12\pi)\ln L$, where L must be proportional to R. The second derivative of this term thus agrees with the difference found just above. Making the correction to (9) then gives

$$\delta F = -\frac{c}{24}\left(\frac{\phi}{\pi} + \frac{\pi}{\phi}\right)\frac{\delta R}{R}, \tag{10}$$

whereby (7) and (10) are consistent.

Integrating (7) from $\phi_0$ to $\phi$, integrating (10) from $r_c$ to R, and equating the results leads to the preliminary result

$$F = -\frac{c}{24}\left(\frac{\phi}{\pi} + \frac{\pi}{\phi}\right)\ln(R/r_c) - \frac{c}{8}\ln\phi. \tag{11}$$

In arriving at (11) an integration constant that may be boundary condition dependent has been omitted, as mentioned.

Now we consider the logarithmic terms in $F$, which, as we will see, are not all included in (11). These terms are always proportional to lnL, where L is a characteristic length. Their coefficients are specified, and in the present case arise from the angle $\phi$, the two right angles, and the curved boundary of the circular section. Making use of the formulas in [3] (cf. also [4]) one finds that these are, respectively, $(c/24)(\phi/\pi-\pi/\phi)$, $(c/24)(-3)$, and $(c/24)(-2\phi/\pi)$. Taking the sum, we see that (11) already contains all these contributions except the term arising from the right angle corners. Since no other logarithmic terms remain, we can simply add the missing term to (11) with the final result

$$F = -\frac{c}{24}\left(\frac{\phi}{\pi} + \frac{\pi}{\phi}\right)\ln(R/r_c) - \frac{c}{8}\ln(S/r_c), \tag{12}$$

where we have introduced the arc length $S = \phi R$. Note that (12) is entirely composed of terms that scale logarithmically with a dilatation.

One might question our choice of L = R in the logarithmic term added above. In fact the most general form for L in the present geometry is $(R/r_c)f(\phi)$, where f is an arbitrary function. But an $f \neq 1$ would result in a new scale invariant term in $F$ with a non-vanishing f derivative, that should already have been included in (11).

Some consequences of (12) have been discussed elsewhere [12]. Further discussion is included in Section V below.

## III. Finite Helical Staircase

In this section we consider a related geometry, the helical staircase of finite angular extent. In polar coordinates this is the region ($a \leq r \leq R$, $0 \leq \theta \leq \phi$), so this geometry may be thought of as a circular section with the inner part removed. $F$ for an infinite staircase has been discussed elsewhere [3].

Our procedure for calculating the free energy is in outline the same as for the circular section. We make use of (6) and (8) to calculate the derivatives of $F$ for an angular and radial stretch, respectively. However, in order to calculate the stress tensor T and evaluate the integrals, we employ a transformation of the helical region to a rectangle. The integrals then include terms proportional to derivatives of the rectangle free energy, already evaluated in [4]. The resulting expressions, including a logarithmic term due to the curved boundaries, are then integrated with respect to (inner) radius and angle, and the unspecified integration functions evaluated by comparison with each other and the rectangle, in an appropriate limit.

The transformation

$$\zeta = R \cdot \exp\left(i \frac{\ln(R/a)}{L'} w\right) \tag{13}$$

maps a rectangle in the w plane ($w = u + iv$) of height L' and width L into a helical region in the $\zeta$ plane ($\zeta = r\, e^{i\alpha}$) of angle

$$\phi = \left(\frac{L}{L'}\right) \ln\left(\frac{R}{a}\right). \tag{14}$$

In performing the calculations it is useful to set L' = K'(k) and L = 2K(k), where K and K' are complete elliptic integrals. Note that a line segment of fixed radius (angle) in the helical region corresponds to a horizontal (vertical) line in the rectangle.

The expectation values of the corresponding stress tensors are found to be connected by

$$\langle T(\zeta)\rangle \zeta^2 = -\left(\frac{K'}{\ln(R/a)}\right)^2 \langle T(w)\rangle + \frac{c}{24}. \tag{15}$$

The necessary stress tensor component is then seen to be

$$\langle T_{rr}\rangle r^2 = -\left(\frac{K'}{\ln(R/a)}\right)^2 \langle T(w) + \overline{T}(\overline{w})\rangle + \frac{c}{12}. \tag{16}$$

At this point, it is useful to recall the known result for the free energy of a rectangle of dimensions L by L' [4],

$$F_r = -\frac{c}{4}\ln L + F_0,$$
$$F_0 = \frac{c}{2}\ln[\eta(q)], \qquad (17)$$

where $\eta$ is the Dedekind function $\eta(q) = q^{\frac{1}{24}}\prod_{n=1}^{\infty}(1-q^n)$, $q = \exp(-2\pi x)$, and the aspect ratio $x = L'/L$.

The radial stretch we employ in the helical region changes the inner radius, $r \to r + r\,\delta a/a\,\Theta(r_0-r)$, $\alpha \to \alpha$. This transformation leads to two terms, one being an integral along the horizontal direction of the rectangle, which, as mentioned, is expressible as a derivative of its free energy. The helical free energy derivative is then

$$\delta F_1 = \left(\frac{\partial F_0}{\partial v}\right)\delta v - \frac{c}{24\pi}\phi\frac{\delta a}{a}, \qquad (18)$$

where w is to be understood as a function of $\zeta$, and the overall expression is denoted $F_1$, since it does not yet contain the logarithmic term from the curved boundaries. This latter contribution is easily seen to be $-c\phi/12\pi\,\ln(R/a)$. Including this term and integrating the result with respect to a, one finds

$$F = \frac{c}{2}\ln\eta(q(y)) - \frac{c\phi}{24\pi}\ln\left(\frac{R}{a}\right) + g(R,\phi), \qquad (19)$$

where the variable $y = \ln(R/a)/\phi$, and the function g is yet to be determined.

The first term on the r. h. s. of (18) (and similarly the first term on the r. h. s. of (20) below) assumes a particularly simple form since the changes of free energy and the stretching transformations in the two geometries are directly mapped onto each other.

The calculation for an azimuthal stretch is very similar to the foregoing, except that one arrives at an integral along the vertical direction of the rectangle. The result is

$$\delta F = \left(\frac{\partial F_0}{\partial u}\right)\delta u + \frac{c}{24\pi}\ln(R/a)\cdot\delta\phi. \qquad (20)$$

After integration over $\phi$ a second expression for the free energy of the helical region results,

$$F = \frac{c}{2}\ln\eta\left(q\left(\frac{1}{y}\right)\right) - \frac{c\phi}{24\pi}\ln\left(\frac{R}{a}\right) + h(R,a). \qquad (21)$$

Making use of the modular property $\ln\eta(1/x) = (1/2)\ln x + \ln\eta(x)$ one finds from comparison of (18) and (19)

$$F = \frac{c}{2}\ln\eta(q(y)) - \frac{c\phi}{24\pi}\ln\left(\frac{R}{a}\right) - \frac{c}{4}\ln\phi + g(R). \qquad (22)$$

In order to evaluate the as yet undetermined function g(R) we take the limit a -> R, φ -> 0, with L'/L held fixed. Thus y = ln(R/a)/φ -> L'/L and the helical region becomes a (vanishingly small) rectangle. In this limit (17) and (22) must coincide, which imposes g(R) = - (c/4) lnR. Introducing the variable of x = 1/y to correspond with earlier expressions then gives rise to the final form

$$F = \frac{c}{2}\ln\eta(q(x)) - \frac{c\phi}{24\pi}\ln(\frac{R}{a}) - \frac{c}{4}\ln\left[R\ln\frac{R}{a}\right]$$

$$x = \frac{\phi}{\ln\frac{R}{a}}.$$

(23)

Note that (23) includes a term $-c/4 \ln R$ which fully accounts for the logarithmic contribution of the four corners. The contribution of the curved boundaries was explicitly added to (19), so (23) gives in the limit φ -> ∞, a result in agreement with the infinite helical staircase geometry ( equation (2.11) of [3]),

$$F/\phi = -\frac{c}{24}\left(\frac{\ln(R/a)}{\pi} + \frac{\pi}{\ln(R/a)}\right).$$

That the free energy of a rectangle is correctly reproduced has already been built into the derivation of (23). The same calculation also describes the limit a, R -> ∞, φ -> 0 with L and L' fixed.

Finally, we may compare (23) with the result (12) for a circular section by letting a -> $r_c$, where $r_c$ is small. In this limit x -> 0, so that $\ln\eta(q(x))$ -> $-\pi/12x - 1/2 \ln x$. Substituting these limiting forms in (23) gives rise to an expression that, up to a constant, is exactly the same as (12), except that the coefficient of the second term on the r. h. s. is now c/4 rather than c/8. However this difference is exactly what one expects, given that the helical region includes four right angle corners while the circular section has only two. Further discussion of $F$ for the staircase is contained in Section V.

**IV. Matrix Elements of the Transfer Matrix**

The result (17) for the rectangle free energy has some implications for certain matrix elements of the transfer matrix in a strip. Similarly, one obtains information about matrix elements of the corner transfer matrix by consideration of the helical region free energy (23). In previous work, Cardy has found similar results for a strip periodic in one direction [13].

We proceed by taking advantage of the connection between the infinitesimal transfer matrix per unit length $\hat{H}_s$ in a strip of width L and $\hat{L}_0$, the generator of scale transformations in the half-plane [3], and a similar connection between $\hat{L}_0$ and $\hat{H}_c$, the infinitesimal corner transfer matrix per unit angle on the helical staircase.

Making use of Equation (3.28) [11] and (2.9) [3], one finds

$$\hat{H}_s = \frac{\pi}{L}\left(\hat{L}_0 - \frac{c}{24}\right),$$

$$\hat{H}_c = 2\epsilon\hat{L}_0 - c\left(\frac{\epsilon}{12} + \frac{1}{48\epsilon}\right),$$

(24)

where

$$\varepsilon = \frac{\pi}{2\ln(R/a)}. \tag{25}$$

Thus

$$\hat{H}_c = 2\varepsilon \frac{L}{\pi} \hat{H}_s - \frac{c}{48\varepsilon}. \tag{26}$$

Now one can calculate the free energy using a transfer matrix running in the L direction, making use of (24). The result is

$$F_r = -\ln\left(\sum_n |<n|B>|^2 q^{n/2}\right) - \frac{\pi c}{24} x, \tag{27}$$

where |B> is the boundary state at the end of the rectangle and |n> are $\hat{L}_0$ eigenstates, with $\hat{H}_s$ energies $n\pi/L$ above the ground state.

Comparing (17) and (27), one finds

$$\sum_n |\langle n|B\rangle|^2 q^{n/2} = L^{c/4}\left(\prod_n \frac{1}{1-q^n}\right)^{c/2}. \tag{28}$$

Now the r. h. s. has an expansion in (whole) powers of q. Thus only even values of n contribute in the sum. This implies a matrix element selection rule $M_n = |<n|B>|^2 = 0$ for n odd. Further, one may determine the values of the $M_n$ as a function of L and c, e.g. $M_0 = L^{c/4}$, $M_2 = c/2\, L^{c/4}$, etc. Since

$$\prod_n \frac{1}{1-q^n} = \sum_n P(n)q^n, \tag{29}$$

where P(n) are the partitions of n, one may in fact express the $M_n$ in terms of these quantities. Note that (29) implies that all $M_n$ contain a universal factor $L^{c/4}$ which arises from the logarithmic corner contribution to the free energy. Such a factor does not appear in the periodic cylinder geometry [13]. Note finally that our procedure omits a possible overall multiplicative constant that depends on the boundary state, as encountered in *F* above.

The free energy of the helical region may be written by use of the corner transfer matrix formalism. By an analogous procedure, employing (23), one finds

$$\sum_n |\langle n|R\rangle|^2 q^{n/2} = \left(R\ln\frac{R}{a}\right)^{c/4}\left(\prod_n \frac{1}{1-q^n}\right)^{c/2}, \tag{30}$$

where |R> is the corner transfer matrix boundary state, along the straight edges of the helical region. The conclusions are the same as for the matrix elements of the (ordinary) transfer matrix, except that L is replaced by R ln(R/a).

## V. Total Free Energies of the Circular Section and Staircase

In this Section we examine some consequences of the Casimir terms (12) and (23) for the total free energy of the circular section and finite helical staircase, respectively. Our analysis follows the lines of the discussion of the rectangle free energy in [9]. To begin, it is useful to recall the results of that work. For an L by L' rectangle of aspect ratio $x_r = L'/L$ and area $A = LL'$, the total free energy is given by

$$\mathbf{F}_r = f_0 A + 2f_1 \sqrt{A}\left(\sqrt{x_r} + \frac{1}{\sqrt{x_r}}\right) + \frac{c}{4}\ln\left[\eta(x_r)\eta\left(\frac{1}{x_r}\right)\right] - \frac{c}{8}\ln A, \qquad (31)$$

where the non-universal constants $f_0$ and $f_1$ are the free energy per unit area (or bulk free energy) and free energy per unit edge length, respectively, the edge length $E = 2(L+L')$ has been expressed in terms of A and $x_r$, and the modular relation for $\eta$ mentioned above has been employed to re-express (17). We use the constant A constraint since it is appropriate in many physical and model situations, for instance when the system consists of a fixed number of molecules at given (surface) density.

It is useful to approximate the Casimir term in (31) via [4]

$$\frac{c}{4}\ln\left[\eta(x_r)\eta\left(\frac{1}{x_r}\right)\right] \cong -\frac{c\pi}{24}\left(x + \frac{1}{x}\right), \qquad (32)$$

which is correct for large or small x, a reasonable approximation otherwise [4], and allows a simple understanding of the effects of the Casimir term in this context. It is then clear that (for $c > 0$) $\mathbf{F}_r \rightarrow -\infty$ as $x_r \rightarrow 0$ or $x_r \rightarrow \infty$. This behavior is due to the attraction of the walls of the rectangle, which is proportional to their length (and inversely proportional to their separation). The leading term is the same as for an infinite strip [2]. Now the edge free energy (second term on the r.h.s. of (31)) behaves oppositely, with a minimum at $x_r = 1$ (the square), since $f_1 > 0$. Furthermore this term grows with area. Thus the edge free energy and Casimir terms compete with each other. This is typical in many geometries, as we will see. The consequence here is that for A sufficiently large, there is a minimum of $\mathbf{F}_r$ at $x_r = 1$, while when A is less than the cross-over area $A_c$, the system exhibits a *Casimir instability*, with no minimum of $\mathbf{F}_r$ at any finite aspect ratio. Then the shape of the rectangle at equilibrium is *elongated* [9]. The cross-over occurs at the area for which the second derivative of $\mathbf{F}_r$ at $x_r = 1$ vanishes. Using (32) gives

$$A_c = \gamma\left(\frac{c}{f_1}\right)^2. \qquad (33)$$

where $\gamma = (\pi/12)^2 = 6.85 \; 10^{-2}$.

The situation is in fact slightly more complicated. There is never a lower bound on $\mathbf{F}_r$, even when $A > A_c$. However this does not imply an instability to large or small $x_r$ at all A, since as A increases above $A_c$ the minimum broadens so much that the relevant $x_r$ values are unphysical, the system being less than one atom wide. Thus there is also a range of A values just above $A_c$ where squares and elongated rectangles may be in thermal equilibrium.

We now describe the behavior of the circular section total free energy. It shows some similarity to the rectangle, but is more complicated. The most significant difference is a much stronger dependence of the cross-over area on the ratio $c/f_1$, as we will see.

Making use of (12) and the edge length for a circular section ("pie slice") gives

$$\mathbf{F}_{ps} = 2f_1\sqrt{A}\left(\sqrt{\frac{\phi}{2}} + \sqrt{\frac{2}{\phi}}\right) - \frac{c}{48}\left(\frac{\phi}{\pi} + \frac{\pi}{\phi}\right)\ln\left(\frac{2A}{\phi}\right) - \frac{c}{16}\ln(2\phi A), \qquad (34)$$

where we have expressed the radius R in terms of the area A and angle $\phi$ of the circular section, the reduced area $A = A/r_c^2$, $f_1 = f_1 r_c$, and any non-universal effects of the boundary curvature have been ignored.

$\mathbf{F}_{ps}$ is a smooth function of $\phi$, however unlike the rectangle it is not symmetric. We have investigated it graphically for $0.01 < c/f_1 < 1000$. For a given value of $c/f_1$, its form varies in a characteristic way with $A$. For large areas, the non-universal term dominates. Then there is a minimum at an angle that approaches $\phi = 2$, where the edge length E is minimal, as $A \to \infty$. This is the leading finite-size behavior in the thermodynamic limit. For small angles $\phi$, however, the Casimir term always dominates so that, for $c > 0$, $\mathbf{F}_{ps}$ approaches $-\infty$ as $\phi \to 0$. This decrease in free energy occurs at increasingly small angles as the area grows, so its onset will occur at an unphysically small angle for large enough area, depending on the system. Returning to the minimum near $\phi = 2$, we find that as $A$ decreases, it becomes more shallow. This is due to the increasing influence of the Casimir term, which has a maximum in this region (as long as $A > 2.25$). The minimum in $\mathbf{F}_{ps}$ finally disappears at a cross-over area $A_c$, where it becomes monotonically increasing with $\phi$. Thus the lowest total free energy occurs at $\phi = 0$. A typical case is shown in Fig. 1. The resulting *Casimir instability* to small angles, or sharp corners, may be regarded as due to the attraction of walls, as discussed below. It occurs if $c/f_1$ is about 15 (for which $A_c = 25$) or larger. For $c/f_1$ in the range from 15 to 50, the cross-over area $A_c$ is reasonably well approximated by (33), but with $\gamma = 0.38$, 5.5 times the value for the rectangle. For $c/f_1$ in the range 50 to 400, $\gamma = 1.09$ is appropriate, which is an increase of a factor 15.9. Thus the *Casimir instability* sets in at much larger areas than for a rectangle, and grows more rapidly with $c/f_1$.

It is of interest to estimate the strength of the Casimir instability. Consider the case illustrated in Fig. 1. Assume, for definiteness, that the angle is constrained to vary between a fixed minimum $\phi_{min}$ and $2\pi$. Then the difference in free energy $\Delta\mathbf{F}_{ps}/c$ between $\phi = 2\pi$ and $\phi = \phi_{min}$ is one measure of the strength. For $\phi_{min} = 0.5$, $\Delta\mathbf{F}_{ps}/c$ increases from 0.316 ($k_B T_c$) at $A = 50$ to a maximum of 0.339 at 158, then decreases to 0.312 at 450. For $\phi_{min} = 0.1$, it decreases from 2.79 to 1.44 as $A$ increases. At the cross-over area $A_c = 147$, it is 2.48.

Returning to the general behavior of $\mathbf{F}_{ps}$, we find that as the area drops further below $A_c$, the Casimir term becomes yet more dominant, and a maximum in $\mathbf{F}_{ps}$ reappears. For small values of $c/f_1$, it is accompanied by a new minimum between it and $\phi = 2\pi$. This feature is seen, for instance, at $A = 32$ for $c/f_1 = 80$, with the minimum at about $\phi = 5.4$. For larger $c/f_1$ values, $\mathbf{F}_{ps}$ decreases for $\phi$ above the maximum so that $\phi = 2\pi$ is a (local) minimum.

We have seen that the *Casimir instability* for the circular section creates a thermodynamic force favoring a small angle, or sharp corner. This effect arises from the $-c\pi/24 \ln(R/r_c)/\phi$ term on the r.h.s. of (12). For small angles, this term can be derived quite simply by integrating $-c\pi/24 \, 1/L$, the free energy per unit length of an infinite strip of width L [2], with $L = \phi r$. Thus the instability, or tendency to sharpen, may be regarded as an effect of the attraction of walls. It may therefore be expected to occur quite generally. For the staircase geometry, as discussed below, we exhibit it explicitly.

The total free energy of the finite staircase is a complicated function because it depends on three parameters. It is not clear that a comprehensive and generally applicable analysis is possible, or of use. We therefore restrict ourselves to the investigation of a few limits and a comparison with the rectangle for the case of small curvature.

The first, and perhaps most interesting case is the limit for which the inner radius a -> 0, when the staircase approaches the circular section. This was discussed briefly at the end of Section III. In this case it follows from (23) that

$$F_{st} \to -\frac{c}{24}\left(\frac{\pi}{\phi}+\frac{\phi}{\pi}\right)\ln\left(\frac{R}{a}\right)-\frac{c}{4}\ln(R\phi), \qquad (35)$$

where we have kept all diverging and finite terms. Thus $F$ (for $c > 0$) becomes arbitrarily negative. What makes this limit interesting is that the instability occurs at *any* area A *without* the edge length E diverging. Therefore the total free energy exhibits a new type of *Casimir instability* in this limit. It is not necessary for the system to be small or become elongated. Since our argument given above for the circular section applies here, this instability is attributable to an attraction of the edges, at least for small angles. Thus it is to be expected in any critical system for which sharp corners can form on the edge.

Introducing $\lambda = R/a$ and using the staircase area A to expresses a in terms of $\phi$ and $\lambda$, one finds

$$\begin{aligned}
\mathbf{F}_{st} &= f_0 A + f_1 E + F_{st}, \\
F_{st} &= \frac{c}{2}\ln\eta(x_{st}) - \frac{c\phi}{24\pi}\ln\lambda + \\
&\quad \frac{c}{8}\ln\left[\frac{\lambda^2-1}{(\lambda\ln\lambda)^2}\right] - \frac{c}{8}\ln\left(\frac{2A}{\phi}\right), \\
E &= 2\sqrt{A}\left(\sqrt{x_r}+\frac{1}{\sqrt{x_r}}\right),
\end{aligned} \qquad (36)$$

where $x_r = \frac{\phi}{2}\left(\frac{\lambda+1}{\lambda-1}\right)$ is the aspect ratio of a rectangle with the same area A and edge length E as the staircase, and $x_{st} = \frac{\phi}{\ln\lambda}$. The form (36) is useful for investigating the properties of the staircase total free energy at constant area. Note that again, any non-universal effects of the boundary curvature have been ignored.

If we let $\phi \to \infty$ and $\lambda \to 1$, keeping a constant, the geometry becomes that of a long, thin ribbon with edges of arbitrary curvature. Setting $\lambda = 1+\delta$, the leading two terms in the total free energy are found to be

$$\mathbf{F}_{st} \to \frac{A}{a^2}\left(2f_1 a\frac{1}{\delta} - \frac{c\pi}{24}\frac{1}{\delta^2}\right), \qquad (37)$$

which diverges to $-\infty$ as $\delta \to 0$. This effect is due to the increase in Casimir attraction of the sides, which become longer and closer together. Each of these effects contributes a divergence proportional to $1/\delta$ to the second term on the r.h.s of (37).

Equation (37) is interesting because the two terms on the r.h.s. of (37) have the same area dependence, so the change from edge dominated to Casimir behavior comes at a $\delta$ value independent of A. This is not the usual behavior, as we have seen for the rectangle and circular section, and may be observed also for many other limits of the staircase total free energy (36) (the a -> 0 limit investigated above is another exception). The origin of the effect here may be understood by regarding the geometry as a rectangle of length $L' = a\phi = A/a\delta$ and width $L = a\delta$. Then the leading Casimir term agrees with (37) and the edge free energy reproduces the non-universal contribution.

The $a \to 0$ instability in (35) provides an illustration of the complications of $\mathbf{F}_{st}$. One might expect that setting $x_r = 1$, for which the edge length E is minimal, is a good starting point for an analysis, since this is the value favored in the thermodynamic limit. However this condition only specifies a line in the $\phi$, $\lambda$ plane. At one end of the line ($\phi \to 0$) $\mathbf{F}_{st}$ approaches a constant, while at the other ($\phi \to 2$) (35) applies, so that $\mathbf{F}_{st}$ diverges to $-\infty$. At any given point along the line there will be a change from Casimir to edge dominated behavior as the area grows, however the area at which this occurs will depend on the point.

Finally, we consider the case of small curvature, to contrast the staircase and rectangle geometry. Unfortunately, for fixed area this is not a well-posed problem. Due to the Casimir term, for any staircase of given A one can find an aspect ratio $x_r$ such that the corresponding rectangle has a smaller total free energy, and conversely for any given rectangle one can find a staircase with lower total free energy. Therefore, in order to be able to compare the two geometries, we impose a constraint of fixed edge length E. We are not aware of any physical situation requiring this condition, but it has the advantage of allowing a comparison of the free energies. Then we let $\lambda = 1 + \delta \to 1$ and $\phi \to 0$ keeping $x_r$ fixed. Thus a, R $\to \infty$. This limit resembles the one employed in deriving (23), except that here A is constant. It follows that, to leading order,

$$\mathbf{F}_{st} = \mathbf{F}_r - \frac{c}{8}\delta. \qquad (38)$$

Thus, at constant A and E (when $c > 0$), the rectangle is unstable to small equal curvature of two opposite sides (ignoring, as above, any non-universal effects of the boundary curvature).

## VI. Conclusions

By use of conformal field theory, we have calculated universal terms for the free energy at criticality in two planar geometries, a circular section ("pie slice") and a finite helical staircase. Some consequences for certain matrix elements of the transfer matrix and corner transfer matrix are derived. We have also examined the total free energy, including non–universal edge free energy terms (but ignoring any non-universal boundary curvature effects), in both cases. For the circular section, a new form of *Casimir instability*, favoring small angles (sharp corners) is found, and we show that the onset of this effect comes at significantly larger areas than in the rectangular geometry. The staircase is shown to have a similar instability to small angle. In this geometry, however, the edge length remains fixed, so there is no competition with the edge free energy term. Arguments are given that this behavior is quite general. Another limit of the staircase is investigated, for which the competition between edge and Casimir terms is independent of area, by contrast to the usual situation. Finally, we show that at constant area and edge length, the rectangle is unstable against small curvature.

## Acknowledgments

We acknowledge a useful conversation with J. L. Cardy. PK is grateful for support from the Deutsche Forschungsgemeinschaft and the Freie Universität Berlin.

**Figure Caption**

Figure 1: Free energy of the circular section ($F_{ps}/c$, see text) versus angle $\phi$ for $c/f_1 = 25$ and, reading from the lowest curve upward, reduced area $A = 50, 150, 300, 450$, respectively.

Free Energy

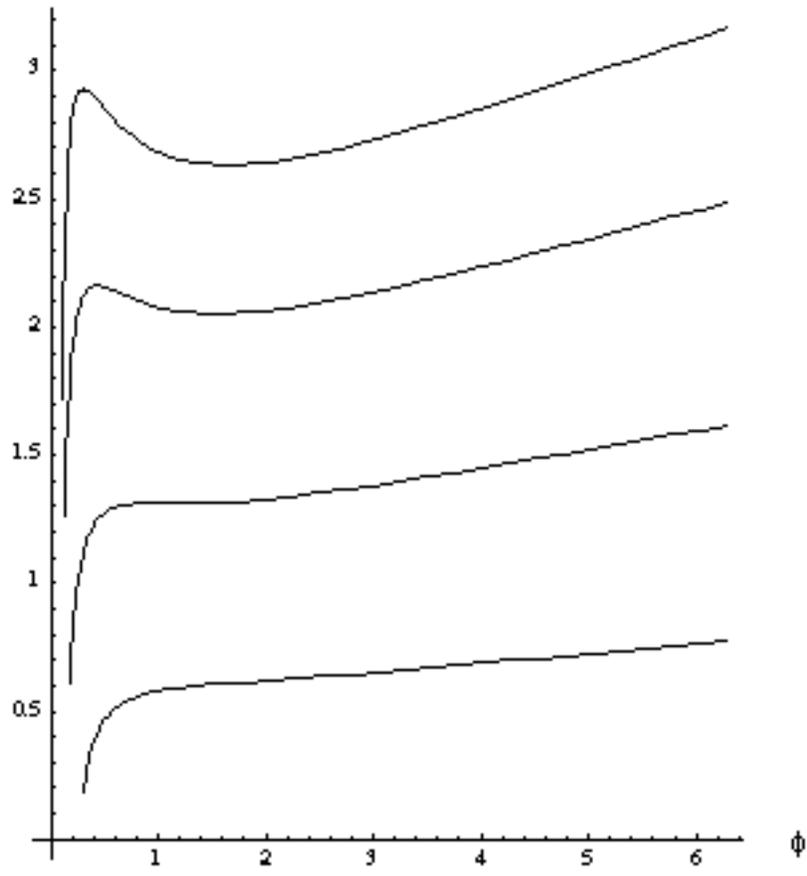